\documentclass[aps,notitlepage]{revtex4-1}

 \usepackage{graphicx}
 \usepackage{epstopdf}
 \usepackage{color}
 \usepackage[colorlinks=true, urlcolor=navyblue, linkcolor=navyblue, citecolor=navyblue]{hyperref}
 \usepackage{lineno}
 \usepackage[normalem]{ulem}

\definecolor{navyblue}{rgb}{0,0.08,0.45}
\definecolor{darkred}{rgb}{0.7,0.0,0.0}
\definecolor{darkgreen}{rgb}{0,0.6,0.2}

\newcommand{\beq}{\begin{equation}}
\newcommand{\enq}{\end{equation}}
\newcommand{\beqa}{\begin{eqnarray}}
\newcommand{\beqast}{\begin{eqnarray*}}
\newcommand{\enqa}{\end{eqnarray}}
\newcommand{\enqast}{\end{eqnarray*}}
\newcommand{\nn}{\nonumber}
\newcommand{\bec}{\begin{center}}
\newcommand{\enc}{\end{center}}
\newcommand{\beqo}{\begin{quote}}
\newcommand{\enqo}{\end{quote}}
\newcommand{\bem}{\begin{minipage}}
\newcommand{\enm}{\end{minipage}}

\newcommand{\lb}{\label}

\newcommand{\req}[1]{(\ref{#1})}
\newcommand{\pa}{\partial}
\newcommand{\half}{\textstyle \frac{1}{2}}

\newcommand{\ze}{\zeta}

\newcommand{\ka}{\kappa}
\newcommand{\la}{\lambda}

\newcommand{\ph}{\phi}

\newcommand{\om}{\omega}

\newcommand{\De}{\Delta}

\definecolor{darkgreen}{rgb}{0,.5,0}
\definecolor{navyblue}{rgb}{0,0.08,0.45}
\definecolor{darkred}{rgb}{0.7,0.0,0.0}
\definecolor{darkgreen}{rgb}{0,0.6,0.2}

\begin{document}

%\linenumbers

%\preprint{SLAC--PUB--16257}

\title{Supersymmetry in the Double-Heavy Hadronic Spectrum}

\author{Marina Nielsen}
\affiliation{Instituto de F\'isica, Universidade de S\~ao Paulo, Rua do Mat\~ao, Travessa R187, 05508-090 S\~ao Paulo, S\~ao Paulo, Brazil}
\affiliation{SLAC National Accelerator Laboratory, Stanford University, Stanford, California 94309, USA \hspace{-8pt} }
\email[]{mnielsen@if.usp.br}

\author{Stanley J. Brodsky}
\affiliation{SLAC National Accelerator Laboratory, Stanford University, Stanford, California 94309, USA}
\email{sjbth@slac.stanford.edu}

\author{Guy F.  de T\'eramond}
\affiliation{Universidad de Costa Rica, 11501 San Pedro de Montes de Oca, Costa Rica}
\email[]{gdt@asterix.crnet.cr}

\author{Hans G\"unter Dosch}
\affiliation{Institut f\"ur Theoretische Physik, Philosophenweg 16, 69120 Heidelberg, Germany}
\email[]{h.g.dosch@thphys.uni-heidelberg.de}

\author{Fernando S. Navarra}
\affiliation{Instituto de F\'isica, Universidade de S\~ao Paulo, Rua do Mat\~ao, Travessa R187, 05508-090 S\~ao Paulo, S\~ao Paulo, Brazil  \hspace{-3pt}} 
\email[]{navarra@if.usp.br}

\author{Liping Zou}
\affiliation{Institute of Modern Physics, Chinese Academy of Sciences, Lanzhou, China}
\email[]{zoulp@impcas.ac.cn}

\date{\today}

\begin{abstract}

Relativistic light-front bound-state equations for double-heavy mesons, baryons and tetraquarks are  constructed  in the framework of supersymmetric light front holographic QCD. Although heavy quark masses strongly break  conformal symmetry, supersymmetry and the holographic embedding of semiclassical light-front dynamics still holds. The theory, derived from five-dimensional anti-de Sitter space,  predicts that the form of the confining potential in the light-front Hamiltonian is harmonic even for heavy quarks. Therefore,  the basic underlying supersymmetric mechanism,  which transforms meson-baryon and baryon-tetraquark wave functions into each other, can also be applied to the double-heavy sector; one can then successfully relate the masses of the double-heavy mesons to double-heavy baryons and tetraquarks. The dependence of the confining potential on the hadron   mass scale agrees completely with the one derived in heavy light systems from heavy quark symmetry. We also make predictions for higher excitations of the charmonium and bottomonium states. In particular, the remarkable equality of the Regge slopes in the orbital angular momentum, $L$, and the principal quantum number, $n$, is predicted to remain valid.

\end{abstract}

%\pacs{11.30.Pb, 12.60.Jv, 12.38.Aw, 11.25.Tq}

\maketitle

\section{Introduction \label{intro}}

Light front holographic QCD (LFHQCD) leads to a remarkable semiclassical approximation to QCD \cite{deTeramond:2008ht,deTeramond:2013it,Brodsky:2014yha}. The basis of LFHQCD is the Maldacena conjecture (or the ``holographic principle") \cite{Maldacena:1997re}, which states the equivalence of a five dimensional classical gravity theory with a four dimensional quantum field theory. The five dimensional classical theory has a non-Euclidean geometry, the so called Anti-de-Sitter (AdS) metric. The resulting four dimensional quantum field theory is a quantum gauge theory, like QCD, but instead of having $N_c = 3$ colours, it has $N_c\to \infty$. It has also conformal symmetry  and it is supersymmetric. This means that to each fermion field there exists also  a bosonic field with properties governed by a supersymmetry.

This superconformal quantum gauge theory with infinitely many colours is not QCD. To consider QCD in LFHQCD one chooses a bottom-up approach: one modifies the five dimensional classical theory in such a way to obtain, from this modified theory and the holographic  embedding, realistic features of hadron physics  which are not apparent in the QCD Lagrangian, such as confinement and the appearance of a mass scale.  In a series of articles~\cite{deTeramond:2014asa,Dosch:2015nwa,Brodsky:2016yod} it was shown how  the implementation of superconformal symmetry~\cite{Akulov:1984uh,Fubini:1984hf,deAlfaro:1976vlx,Haag} of the semiclassical theory, as expressed by holographic light front bound-state equations, completely fixes the necessary modifications  of the AdS$_5$ Lagrangian   for mesons and baryons. Although heavy quark masses break conformal symmetry, the presence of a heavy mass does not need to also break supersymmetry, since it can stem from the dynamics of color confinement. Indeed, as was shown in Refs.~\cite{Dosch:2015bca,Dosch:2016zdv}, supersymmetric relations between the meson and baryon masses still hold to a good approximation even for heavy-light, {\it i.e.}, charm and bottom, hadrons.   In the limit of massless quarks one has an universal scale (fixed for instance by one hadronic mass) and for massive quarks, one has also the quark masses as parameters. This SuSyLFHQCD leads to remarkable relations which connect meson, baryon and tetraquark spectroscopy~\cite{Brodsky:2016yod,Nielsen:2018uyn,Dosch-lec}.

In this work we will show that supersymmetric relations between double-heavy mesons,  baryons, and tetraquarks can still be derived from the supersymmetric algebra even though conformal invariance is explicitly broken by the heavy quark masses. We emphasize that the supersymmetric relations which are derived from supersymmetric quantum mechanics are not based on  supersymmetric Quantum Field Theory  in which QCD is embedded; instead, they are based on the fact that the supercharges of the supersymmetric algebra relate the eigenstates of mesons, baryons and tetraquarks in a Hilbert space in which the light-front (LF) Hamiltonian acts. This could be considered as a realisation of supersymmetric quantum mechanics~\cite{Witten:1981nf}. These relations are possible since in the light-front holographic approach the baryon must be described by the wave function of a quark and a diquark-cluster, and the tetraquark must be  described by the wave function of a diquark-cluster and a antidiquark-cluster. This clustering is purely kinematical, it does not imply  that the diquark cluster  forms a tightly bound system; on the contrary, the results of the form factor analysis \cite{Sufian:2016hwn} show that the cluster is of the usual hadronic size and must be resolved \cite{Brodsky:1983vf}.  The properties of the supercharges predict specific constraints between mesons and baryons, and between baryons and tetraquarks superpartners, in agreement with measurements across the entire hadronic spectrum, including the double-heavy sector~\cite{Nielsen:2018uyn}.

This paper is organized as follows:  In Sec. \ref{SSLFH} we give a brief review of the LF Hamiltonian from supersymmetric quantum mechanics. In Sec. \ref{QRBS} we extend our approach to systems containing double-heavy, charm or bottom, quarks. We compare our predictions with experiment in Sec. \ref{EXP}, and in Sec. \ref{CONCL} we present our conclusions.

\section{Supersymmetric light front Hamiltonian \label{SSLFH}}

In the framework of  supersymmetric quantum mechanics \cite{Witten:1981nf}, the LF Hamiltonian \cite{Dosch:2015nwa,Brodsky:2016yod,Dosch:2015bca,Dosch:2016zdv} can be written in terms of two fermionic generators, the supercharges, $Q$ and $Q^\dagger$, which satisfy anticommutations relations:

\beq
\{Q,Q\} = \{Q^\dagger,Q^\dagger\}=0.
\enq 

The Hamiltonian
\beq 
H=  \{Q,Q^\dagger\} ,
\enq
commutes with these fermionic generators: ${[Q, H]}  = [Q^\dagger, H] = 0$.  Its  minimal realization  in matrix notation is

\beq Q =
\left(\begin{array}{cc}
0&q\\
0&0\\
\end{array}
\right) ,\quad Q^\dagger=\left(\begin{array}{cc}
0&0\\
q^\dagger&0\\
\end{array}
\right) , \label{QQdag}
\enq with 

\beq \label{qdag}
q =-\frac{d}{d\ze} + {f\over\ze}+ V(\ze), \quad  \quad  q^\dagger = \frac{d}{d\ze} +  {f\over\ze}+ V(\ze),
\enq  
where $\ze$ has dimension of length.  The resulting Hamiltonian 
\beq \label{ham} 
H =
\{Q,Q^\dagger\} 
  =     \left(\begin{array}{cc} - \frac{d^2}{d
\ze^2}+\frac{4(f+1/2)^2-1}{4\ze^2}+U_1(\ze)&\hspace{-1cm}0\\
0&\hspace{-1cm} - \frac{d^2}{d \ze^2} +\frac{4 (f-1/2)^2-1}{4\ze^2}+U_2(\ze)
\end{array}\right), \nn
\enq
where
\beqa
U_1(\ze) & = &  V^2(\ze) -V'(\ze) + \frac{2 f}{\ze} V(\ze)    \label{UMa}, \\
U_2(\ze) & = &   V^2(\ze) +V'(\ze) + \frac{2 f}{\ze} V(\ze)  \label{UBa},
\enqa
can be identified with a semiclassical approximation to the QCD LF Hamiltonian of mesons, $H_M=H_{11}$,  and baryons, $H_B=H_{22}$.

In the LFHQCD approach the LF potential is derived from the AdS$_5$ action  from the mapping of the AdS equations to the light front for arbitrary spin~\cite{deTeramond:2013it,Brodsky:2014yha,Dosch-lec}. As has been shown in~\cite{Dosch:2016zdv},  the form of the LF potential \req{UMa} is only compatible with the one derived from an arbitrary dilaton profile in the meson Lagrangian, if 
\beq \lb{linV}
 V(\ze) = \la_Q\,  \ze.
 \enq 
This signifies that, even in the absence of conformal symmetry, the special form of the light front potential for massless quarks persists, provided that the holographic embedding is possible; namely, that  the separation of the dynamical and kinematical variables also persist, at least to a good approximation, in the presence of heavy quark masses~\cite{Dosch:2016zdv}. This can be understood if, to first order, the transverse dynamics is unchanged, and consequently the transverse LF wave function (LFWF) is also unchanged to first order~\cite{deTeramond:2014rsa}. In this case  the confinement  scale  $\la_Q$ takes the place of the confinement scale of massless quarks, $\la = \ka^2$, but depends, however, on the mass of the heavy quark as expected from Heavy Quark Effective Theory (HQET)~\cite{Isgur:1991wq}. The LF potentials, $U_M(\zeta)$ and $U_B(\ze)$, are derived from \req{UMa} and \req{UBa} respectively, 
\beqa
U_M(\ze) & = &  \la_Q^2 \ze^2 + 2 \, \la_Q (f - 1/2),   \label{UM}  \\
U_B(\ze) & = &  \la_Q^2 \ze^2 +  2\, \la_Q (f + 1/2),  \label{UB} 
 \enqa
 with the light-front orbital angular momentum, $L$, for baryons and mesons related by $L_B + \half = L_M - \half \;=f$.

The addition of a constant term to the Hamiltonian \req{ham}  does not violate  supersymmetry and, in the following, we will use  the Hamiltonian~\cite{Brodsky:2016yod}
\beq  \label{Hnu}
H_{S,m_q} =\{Q,Q^\dagger\} + (2 \la_Q S\,+\,\Delta M^2[m_1,...,m_N])\,  {\bf I},
\enq
where $S$ depends on the internal quark spin and the term  $\Delta M^2[m_1,...,m_N]$ is the correction for the quark masses given by Eq.~\req{delm2} in the next section. This term  differs by the additional light quark mass present in baryons or tetraquarks and, therefore, slightly breaks supersymmetry.

The Hamiltonian \req{Hnu} acts on the 4-plet~\cite{Brodsky:2016yod, Nielsen:2018uyn, Dosch-lec}
\beq\label{multi}
|\ph^{LF }\rangle=\left(\begin{array}{cc}
\phi_M{(L_M = L_B+1)} &\psi^{-}{(L_B+ 1)}\\
\psi^{+}{(L_B)} &\phi_T{(L_T = L_B)}
\end{array}
\right),
\enq
with $H_{S,m_q}  |\ph^{LF }\rangle= M^2_{S,m_q}\,  |\ph^{LF }\rangle$.
The resulting expressions for the squared masses of the mesons, baryons and  tetraquarks are~\cite{Brodsky:2016yod,Nielsen:2018uyn}:
\beqa \label{mesfin}  
\mbox{Mesons: } && M_M^2 = 4 \la_Q (n+L_M+{S_M\over2})+ \De M^2[m_1,m_2] ,\\
\label{barfin}\mbox{Baryons: } && M_B^2=4 \la_Q (n+L_B+{S_D\over2}+1) + \De M^2[m_1,m_2,m_3] ,\\
\label{tetrafin}   \mbox{Tetraquarks: } && M_T^2=4 \la_Q (n+L_T+{S_T\over2}+1) + \De M^2[m_1,m_2,m_3,m_4] ,
\enqa 
where $S_M$ is the meson spin, $S_D$ is the lowest possible value of the diquark cluster spin of the  baryons, while $S_T$ is the total tetraquark spin.  The different values of the mass corrections, $\Delta M^2$, on the supermultiplet break supersymmetry explicitly to order  $m_q^2/m_Q^2$,  where we label respectively by $q$ and $Q$ the light and heavy quark masses. These equations show that the excitation spectra of meson, baryon and tetraquark bound states lie on linear Regge trajectories with identical slopes in the radial, $n$, and orbital, $L$, quantum numbers. Mesons with $L_M$ and $S_M$ are the superpartners of baryons with $L_B=L_M-1$ and whose diquark has $S_D=S_M$. Analogously, baryons with $L_B$ and diquark with $S_D$  are the superpartners of tetraquarks with $L_T=L_B$, and $S_T=S_D$. The relation $S_T=S_D$ implies that one of the diquarks in the tetraquark always has spin zero \cite{Nielsen:2018uyn}.

\section{Quarkonium as a relativistic bound state on the light front \label{QRBS}}

A system consisting of two light quarks, or one light and one heavy quark, is relativistic. On the other hand, a system consisting of two heavy quarks is close to the non-relativistic case. However, the front form (FF) formulation (light front dynamics) of the theory of interacting particles is applicable to nonrelativistic as well as relativistic constituents. Therefore, quarkonia can be naturally treated as a relativistic bound state in the LF formulation, as done for instance in~\cite{Branz:2010ub,Li:2017mlw}. In Refs.~\cite{Branz:2010ub,Li:2017mlw} a one gluon exchange interaction, in addition to a hyperfine-splitting contribution~\cite{Branz:2010ub}, or to a longitudinal confining potential~\cite{Li:2017mlw}, was  added to the holographic potential to describe the double-heavy mesonic states. Since it was show in  Ref.~\cite{Trawinski:2014msa}  that a linear confining potential in the instant form of dynamics agrees with a quadratic confining potential in the FF of dynamics, it seems natural to extend the formulation developed in~\cite{Brodsky:2014yha,deTeramond:2014asa,Dosch:2015nwa,Brodsky:2016yod} to a system with two heavy constituents.

In the LF form, the mass for a meson with two massive constituents  in momentum space is given by~\cite{deTeramond:2008ht,Brodsky:2014yha,Dosch-lec}:
\beq M^2= \int_0^1 dx \int d^2 k_\perp \,  \left(   \frac{1}{x(1-x)}
{\vec k_\perp}^2 + \frac{m_1^2}{x}+ \frac{m_2^2}{1-x} \right)
 { \vert \tilde \psi(x, \vec k_\perp) \vert^2 }+ \mbox{interactions}, \lb{lfmk}
\enq
where $\tilde \psi(x, \vec k_\perp)$ is the LFWF of two constituents with relative momentum $\vec k_\perp$ and longitudinal momentum fractions $x_1=x, \; x_2=(1-x)$. For a system with two heavy quarks, $m_1=m_2=m_Q$, Eq.\req{lfmk} can be written as:
\beq 
M^2= \int_0^1 dx \int \frac{d^2 k_\perp}{16 \pi^3} \, \left(   \frac{{\vec k_\perp}^2 +m_Q^2}{x(1-x)}\right)
{ \vert  \tilde \psi(x, \vec k_\perp) \vert^2 }+\mbox{interactions}.\lb{lfmq}
\enq
By using the Fourier transform of ${\psi}(x,\vec b_{\perp})$:
\beq 
\tilde {\psi}(x, \vec k_\perp)={ \sqrt{4\pi}}\int d^2b_{\perp}~e^{-i {\vec k_\perp \cdot} \, \vec b_{\perp}} {\psi}(x,\vec b_{\perp})\lb{fourier},
\enq
in Eq.~\req{lfmq} we obtain:
\beq 
M^2 = \int_0^1 dx \int d^2 b_{\perp}
\, {\psi^*}(x, \vec b_{\perp})  \left( \frac{ - \vec \pa^2_{b_{\perp}} + m_Q^2 }{x(1-x)}\right) {\psi}(x, \vec b_{\perp})+\mbox{interactions},
\enq
with normalization
\beq\label{N}
\int_0^1 \! d x \int  \! d^2 \vec{b}_\perp  \, \left\vert\psi(x, \vec{b}_\perp)\right\vert^2 
= \int_0^1 \! d x \! \int \!  \frac{d^2 \vec{k}_\perp}{16 \pi^3}   \,  \left\vert \tilde\psi (x, \vec{k}_\perp) \right \vert^2  = 1.
\enq

We introduce the invariant impact variable $\zeta = \sqrt{x(1-x)} \vert \vec b_\perp \vert$, which is precisely mapped to the coordinate $z$ of AdS space~\cite{deTeramond:2008ht} by the relation $\zeta = z$. In terms of $\zeta$ we write the LFWF $\psi$ as
\beq \label{psiwf}
\psi(x, \zeta, \varphi) = e^{i L \varphi} \chi(x) \frac{\phi(\zeta)}{\sqrt{2 \pi \zeta}},
\enq
where we have factored out the longitudinal and orbital dependence from the LFWF $\psi$.  From \req{N} the normalization of the transverse and longitudinal modes is given by
\beqa\label{Nphichi}
 \langle\phi\vert\phi\rangle &\!=\!& \int_0^\infty \! d \zeta  \, \phi^2(\zeta) = 1, \\ \label{NL}
 \langle \chi \vert \chi \rangle &\!=\!&  \int_0^1 dx \, { \chi^2(x)\over x(1-x)} =1 .
 \enqa

Using \req{psiwf} we obtain
\beq \label{M2int}
M^2  =  \int \! d\zeta \, \phi^*(\zeta) \sqrt{\zeta}
\left( -\frac{d^2}{d\zeta^2} -\frac{1}{\zeta} \frac{d}{d\zeta}
+ \frac{L^2}{\zeta^2}\right)
\frac{\phi(\zeta)}{\sqrt{\zeta}}
+ \int \! d\zeta \, \phi^*(\zeta) U(\zeta) \phi(\zeta)  + \Delta M_Q^2,
\enq
where
\beq
\Delta M_Q^2 = m_Q^2 \int_0^1 \frac{dx}{x^2(1-x)^2}  \, \chi^2(x),
\enq
The longitudinal function  $\chi(x) \to \sqrt{x~ (1-x)}$ in the limit of zero quark masses~\cite{Brodsky:2006uqa, Brodsky:2014yha}.  

In deriving \req{M2int} we have assumed that separation of transverse and longitudinal dynamics is a good approximation, even in the presence of heavy quark masses, and that the effective  potential, $U$, only depends on the transverse invariant variable $\ze$.

Therefore, also in the case of double-heavy quarks 
and strongly broken conformal invariance, the confinement potential $U$ has the same quadratic form as the one dictated by the conformal algebra. The LF effective transverse potential can still be obtained from holography and is given by Eq.~\req{UM}, at the scale $\lambda_Q$, namely $U(\ze) = \la_Q^2 \ze^2 + 2 \la_Q(L_M - 1)$.
Since the eigenvalues of the LF Hamiltonian
\beq
H = \left( -\frac{d^2}{d\zeta^2} -\frac{1}{\zeta} \frac{d}{d\zeta}
+ \frac{L_M^2}{\zeta^2} + \la_Q^2 \ze^2 + 2 \la_Q(L_M - 1)\right),
\enq
are $4 \la_Q^2 (n + L_M)$, we obtain from \req{M2int}
\beq
M_M^2 =  4 \la_Q^2 (n + L_M) + \Delta M_Q^2.
\enq
for a spinless double-heavy meson. Extension of this result to mesons with internal spin and to baryons and tetraquarks is carried out using the procedures described in Sec. \ref{SSLFH}, if the supersymmetric connection between mesons, baryons and tetraquark bound states holds also for double-heavy quarks.
The masses of the  double-heavy states are  thus given by Eqs.~\req{mesfin},~\req{barfin},~\req{tetrafin}. 

To actually compute $\Delta M$ in \req{M2int} we need to know the longitudinal component of the LFWF $\chi(x)$, which is determined by the holographic mapping only  for massless quarks~\cite{Brodsky:2006uqa}. To this end we follow the procedure introduced in~\cite{Brodsky:2008pg}, in the framework of the holographic soft-wall model~\cite{Karch:2006pv}, for the LFWF of a meson bound state with massive constituents. This procedure amounts to the change
\beq
\frac{{\vec k_\perp}^2}{x(1-x)}\rightarrow \frac{{\vec k_\perp}^2 +m_Q^2}{x(1-x)} ,
\lb{change}
\enq
in the exponential factor in the LFWF in momentum space; the LFWF in impact space then follows from the Fourier transform \req{fourier}.
In particular, for the $n=L=S=0$ meson bound states one obtains for \req{psiwf}~\cite{Brodsky:2014yha, Dosch-lec}
\beq \label{wf} 
{\psi(x, \vec b_\perp)}
= {N_m}\sqrt{\la_Q\over\pi}\sqrt{x(1-x)}\, e^{-{\la_Q\over2} x(1-x)b_\perp^2}\, e^{-{m_Q^2/(2\la_Q x(1-x))}}, 
\enq
with
\beq\lb{norm}
N_m^2={1\over\int_0^1dx~e^{-{m_Q^2/(\la_Q x(1-x))}}},
\enq 
The mass correction, $\Delta M^2$, in \req{mesfin} or \req{M2int} is given by
\beq\lb{delm2}
\De M^2[m_Q,m_Q]=m_Q^2N_m^2\int_0^1dx~{e^{-{m_Q^2/(\la_Q x(1-x))}}\over x(1-x)}.  
\enq

Since we are also interested  in baryons and tetraquarks, which have  more than two constituents,  one has   to form two clusters with $N_a$ constituents each and to introduce the effective $x$ values and transverse separations~\cite{Dosch-lec} : 
\beq
x_a^{\it eff} = \sum_{i=1}^{N_a} x_i , \, \quad \vec b_{\perp,a}^{\it eff} 
=\frac{1}{x_a}\sum_{i=1}^{N_a} x_i \,\vec  b_{\perp,i},  \; \quad a=1,2.
\enq
The resulting light front variable $\zeta$  occurring in the wave function is 
\beq
\ze = \sqrt{x_1^{\it eff}\,x_2^{\it eff}} \left \vert \vec  b_{\perp,1}^{\it eff}- \vec b_{\perp,2}^{\it eff} \right \vert.
\enq
For $n$ constituents the mass correction is then given by~\cite{Brodsky:2016yod}:
\beq
\De M^2[m_1, \cdots ,m_n]= \frac{\la_Q^2}{F} \frac{\rm d F[\la_Q]}{{\rm d} \la_Q} 
\enq
with $F[\la_Q] = \int_0^1 \cdots \int_0^1 e^{-\frac{1}{\la_Q}\sum_{i=1}^{n} m_i^2/x_i}$.

\section{Comparison with experiment \label{EXP}}

\subsection{Mass spectrum}

%%%%%%%%%%%%%%%%%%%%
\vspace{10pt}
\begin{figure}[h] 
\includegraphics[width=8.0cm]{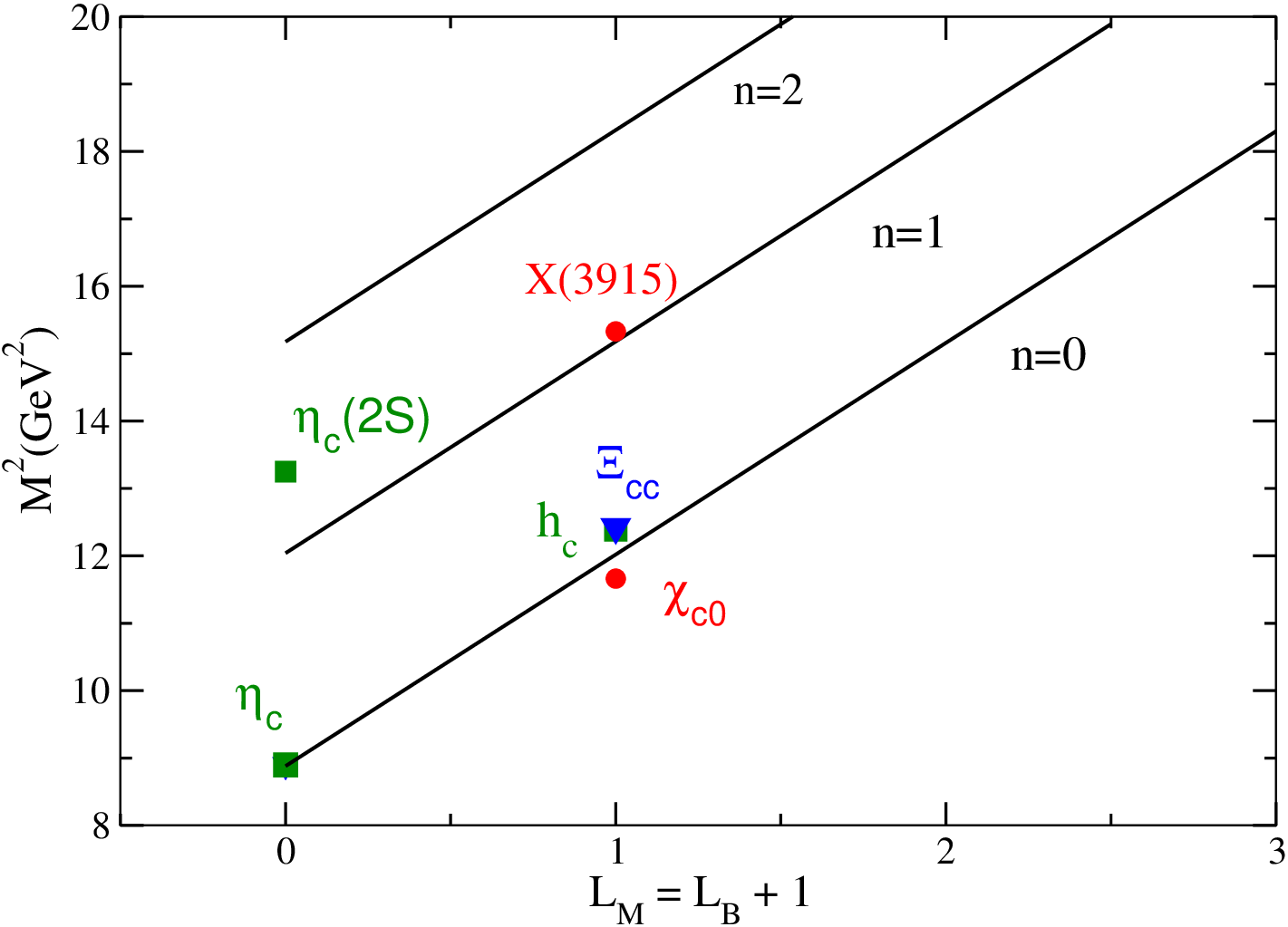}
\hspace{5pt}
\includegraphics[width=8.0cm]{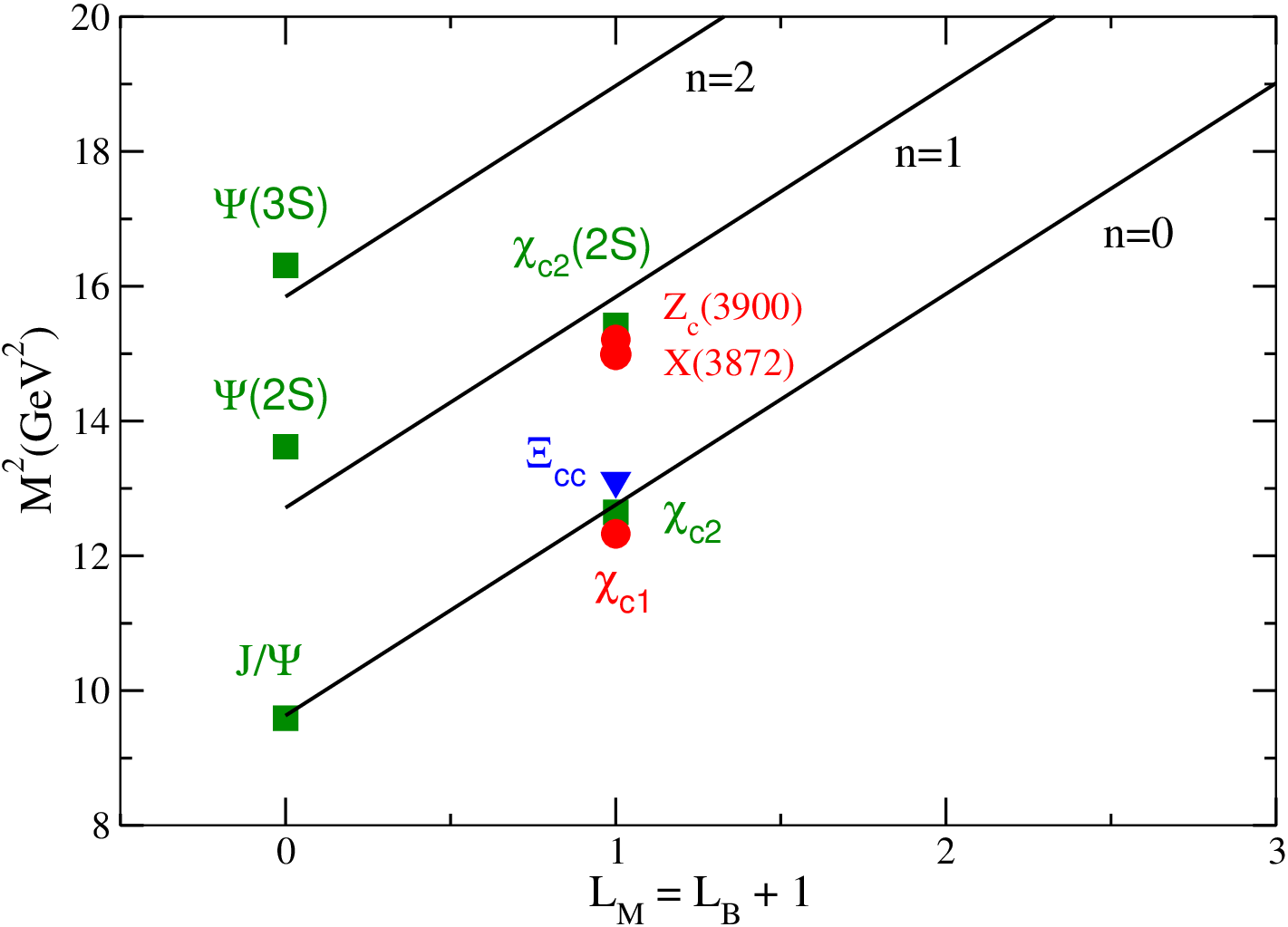}
%\vspace{5pt}
\caption{ \label{charm}  Double charm mesons (shown as green squares) baryons (shown as  blue triangles) and tetraquarks (shown as red circles) with different values of angular momentum $L$ and radial excitation $n$. The solid lines are the trajectories fit from \req{mesfin}. Hadron masses are taken from PDG~\cite{PDG}.  In the left hand side figure we show states with $S_M=S_D=S_T=0$ and we have used $\la_Q=0.785$ GeV$^2$ and $\Delta M^2=8.898$ GeV$^2$ for the values of the parameters in Eq.~\req{mesfin}. In the right hand side  figure we show states with $S_M=S_D=S_T=1$ and we have used $\la_Q=0.782$ GeV$^2$ and $\Delta M^2=8.027$ GeV$^2$ for the values of the parameters in Eq.~\req{mesfin}. }
\end{figure}
%%%%%%%%%%%%%%%%

In Fig.~\ref{charm} we show data~\cite{PDG} for double-charm mesons, baryons and tetraquark superpatners.  The squared masses are plotted against $L_M=L_B+1$; mesons, baryons and tetraquarks with the same abscissa are then predicted to have the same mass. The lightest meson has angular momentum zero and, therefore, does not have a supersymmetric baryon partner~\cite{Dosch:2015bca}. The solid lines in these figures are the trajectories fit from \req{mesfin}. The $L_M=1$ state in the $\eta_c(2984)$ family is the $h_c(3525)$ and the $L_M=1$ state in the $J/\psi(3096)$ family is the $\chi_{c2}(3556)$.  The baryonic superpartner of the meson $h_c(3525)$ is the $\Xi_{cc}$ state with quantum numbers $J^{P}={1\over2}^+$.  There are two candidates for this state: the $\Xi_{cc}^{LHCb}(3620)$ observed in 2017 by the LHCb collaboration \cite{Aaij:2017ueg} and the $\Xi_{cc}^{SELEX}(3520)$ state reported by the SELEX collaboration in 2002 \cite{Mattson:2002vu,Ocherashvili:2004hi}; both masses are well within the uncertainties of our model in the $h_c(3525)$ and $\chi_{c2}(3556)$  mass range.
For additional interpretation concerning the $\Xi_{cc}$ states see Ref.~\cite{Brodsky:2017ntu}. The $\Xi_{cc}^{SELEX}$ and $\Xi_{cc}^{LHCb}$ are the baryonic superpartners of the $h_{c}(3525)$ and $\chi_{c2}(3556)$ mesonic states.
The tetraquark candidates for the  superpartners of the baryonic states $\Xi_{cc}^{SELEX}(3520)$ and $\Xi_{cc}^{LHCb}(3620)$ are the scalar, $J^{PC}=0^{++}$, $\chi_{c0}(3415)$, and the axial, $J^{PC}=1^{++}$, $\chi_{c1}(3510)$, states, respectively, as discussed in \cite{Nielsen:2018uyn}. As tetraquark states, $\chi_{c0}(3415)$ has $L_T=S_T=0$ and $\chi_{c1}(3510)$ has $L_T=0$ and $S_T=1$. As pointed out in Sec.~II, one of the diquarks in the tetraquark has always spin zero. See Ref.~\cite{Nielsen:2018uyn} for more details.

%%%%%%%%%%%%%%%%%
\vspace{17pt}
\begin{figure}[htb] 
\includegraphics[width=8.0cm]{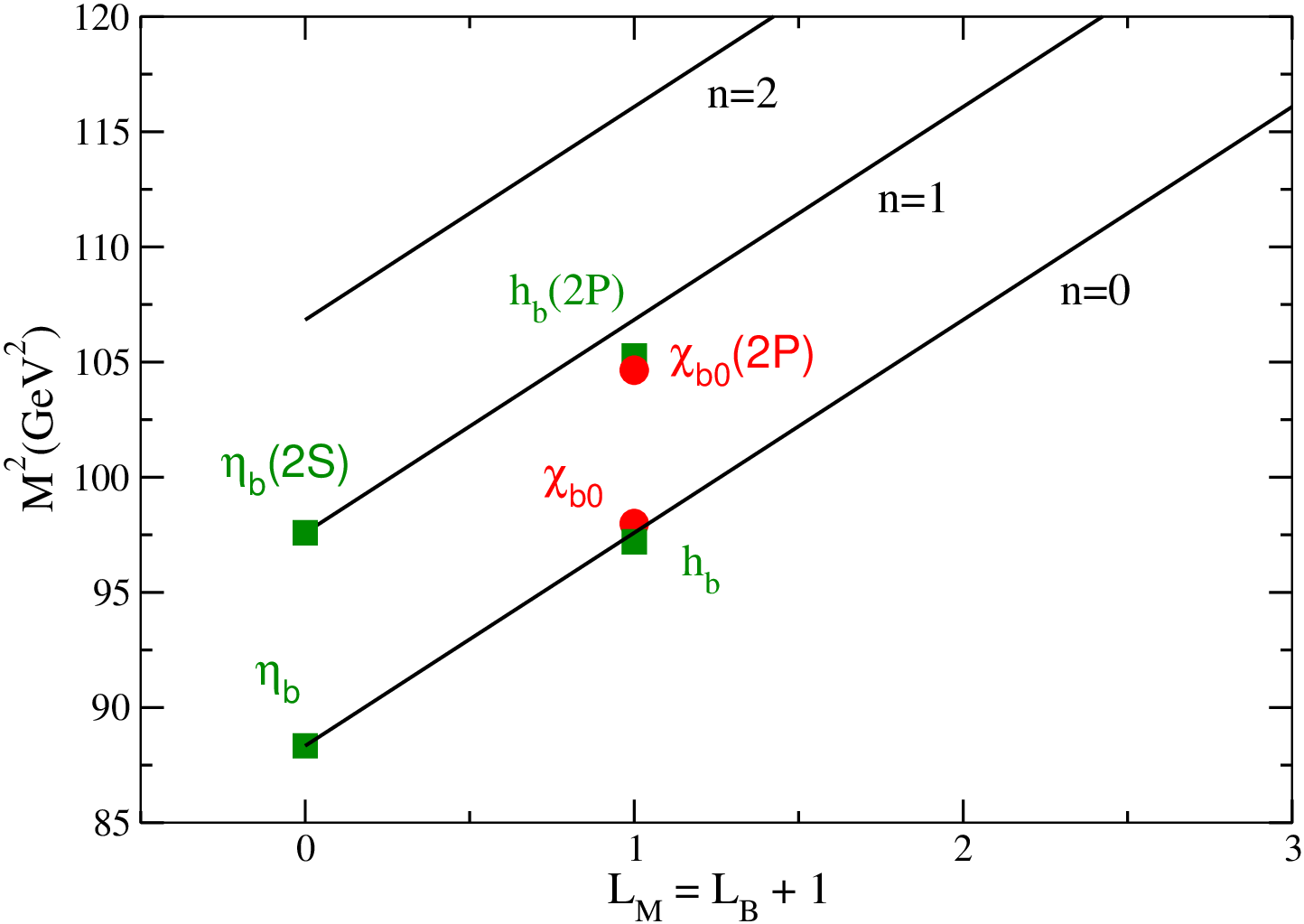}
\hspace{5pt}
\includegraphics[width=8.0cm]{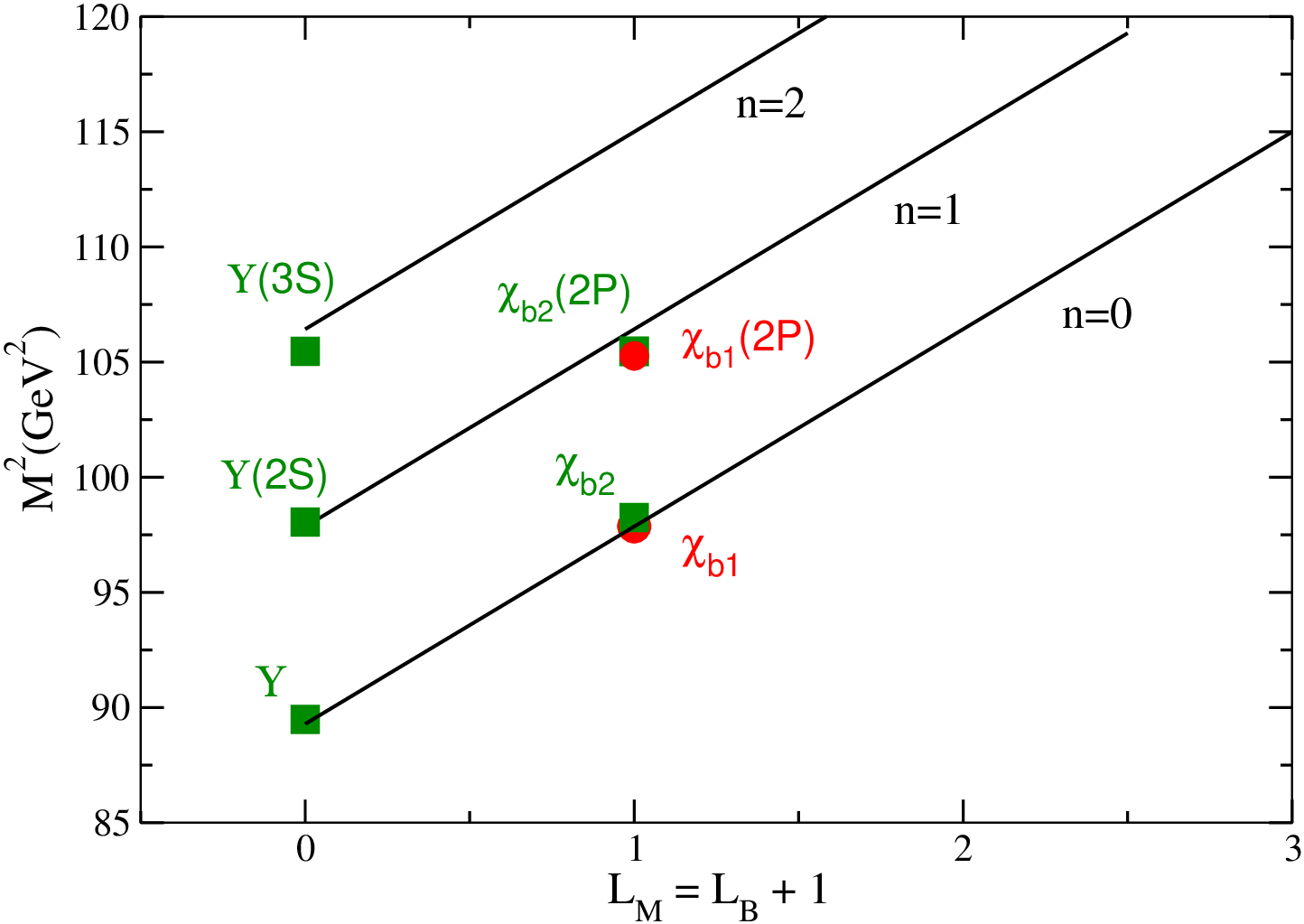}
%\vspace{5pt}
\caption{\label{beauty} Same as in Fig.~\ref{charm} for double beauty hadrons. In the left hand side  figure we show states with $S_M=S_D=S_T=0$ and we have used $\la_Q=2.312$ GeV$^2$ and $\Delta M^2=88.34$ GeV$^2$ for the values of the parameters in Eq.~\req{mesfin}. In the right hand side figure we show states with $S_M=S_D=S_T=1$ and we have used $\la_Q=2.142$ GeV$^2$ and $\Delta M^2=85.01$ GeV$^2$ for the values of the parameters in Eq.~\req{mesfin}.}
\end{figure}
%%%%%%%%%%%%%%%%

In  Fig. \ref{beauty} we show data~\cite{PDG} for double-beauty mesons and tetraquark candidates \cite{Nielsen:2018uyn}. The $L_M=1$ state in the $\eta_b(9400)$ family is the $h_b(9900)$ and the $L_M=1$ state in the $\Upsilon(9460)$ family is the $\chi_{b2}(9910)$. There is still no experimental observation of double beauty baryons. The predicted mass for $\Xi_{bb}$ from this supersymmetric approach is $M_{\Xi_{bb}}=(9.90\pm0.05)$ GeV for both $J^P$ assignments \cite{Nielsen:2018uyn}. The tetraquark candidates for the $J^P={1\over2}^+$ and ${3\over2}^+$ baryonic states are the scalar, $J^{PC}=0^{++}$, $\chi_{b0}(9860)$, and the axial, $J^{PC}=1^{++}$, $\chi_{b1}(9893)$, states respectively.

%%%%%%%%%%%%%%%%%%%
\begin{table}[htb]
\begin{tabular}{|c|c|c|c|}
\hline  
Meson & $M_M$( GeV)  & $\sqrt{\lambda_Q}$( GeV) & Ref. 
\\ \hline 
$\pi$ & 0.14  & $0.57\pm0.03$ & \cite{Dosch:2015nwa}\\
$K$ & 0.50  & $0.57\pm0.03$ & \cite{Dosch:2015bca}\\
$D$ & 1.87  & $0.71\pm0.04$ & \cite{Dosch:2016zdv}\\
$\eta_c$ & 2.98  & $0.90\pm0.04$ & this work\\
$B$ & 5.28  & $1.1\pm0.1$ & \cite{Dosch:2016zdv}\\
$\eta_b$ & 9.40  & $1.49\pm0.03$ & this work
\\ \hline 
\end{tabular} 
\caption{\small \label{lamb-ta}
The fitted value of $\sqrt{\lambda_Q}$ for different meson trajectories as a function of the mass of the lowest meson state on the trajectory. }
\end{table}

Unfortunately the data for double-heavy hadrons are sparse and one cannot really test the predicted linear trajectories. However, the excellent agreement between the masses of the superpartners in the double-heavy-quark sector supports this attempt. From the fits for the different trajectories we arrive at the values for $\sqrt{\lambda_Q}$ shown in Table~\ref{lamb-ta}. In this table we also include the values for $\sqrt{\lambda_Q}$ obtained for light \cite{Dosch:2015nwa}, light-strange \cite{Dosch:2015bca}, one-charm and one-beauty \cite{Dosch:2016zdv} states.

To test the predicted identical linear slopes in the Regee trajectories in the radial, $n$, and orbital, $L$, quantum numbers (see Eqs.~\req{mesfin}, \req{barfin} and \req{tetrafin}), we also show in Figs. \ref{charm} and \ref{beauty} the $n=1$ observed states and the predicted Regge trajectories. As one can see from these figures, the agreement is quite good. As discussed in  \cite{Nielsen:2018uyn}, we can assign the new charmonium states $X(3872)$ and $Z_c^+(3900)$  as natural candidates to the tetraquark superpartners of the   $\chi_{c2}(2P)$ state with an impressive agreement. In the case of the $h_c(2P)$ state, although this state has not been observed yet, the prediction for its mass \cite{Lebed:2016hpi}, is in excelent agreement with the mass of the new charmonium state $X(3915)$, candidate for its tetraquark superpartner. For completeness we also include in these figures the observed states $\psi(3S)$ and $\Upsilon(3S)$ and the predicted Regee trajectory for $n=2$.

In  \cite{Dosch:2016zdv} it was shown that  for heavy-light mesons consistency with  HQET~\cite{Isgur:1991wq} requires that  the confining  scale, $\la_Q$, has for  heavy quark masses, to be proportional to the mass of the heavy meson:
\beq 
\lb{hq} \sqrt{\la_Q}=C \, \sqrt{M_M}, 
\enq
where $C $ is a constant with dimension [mass$^{1/2}$]. 

\begin{figure}[h] 
\includegraphics[width=10.0cm]{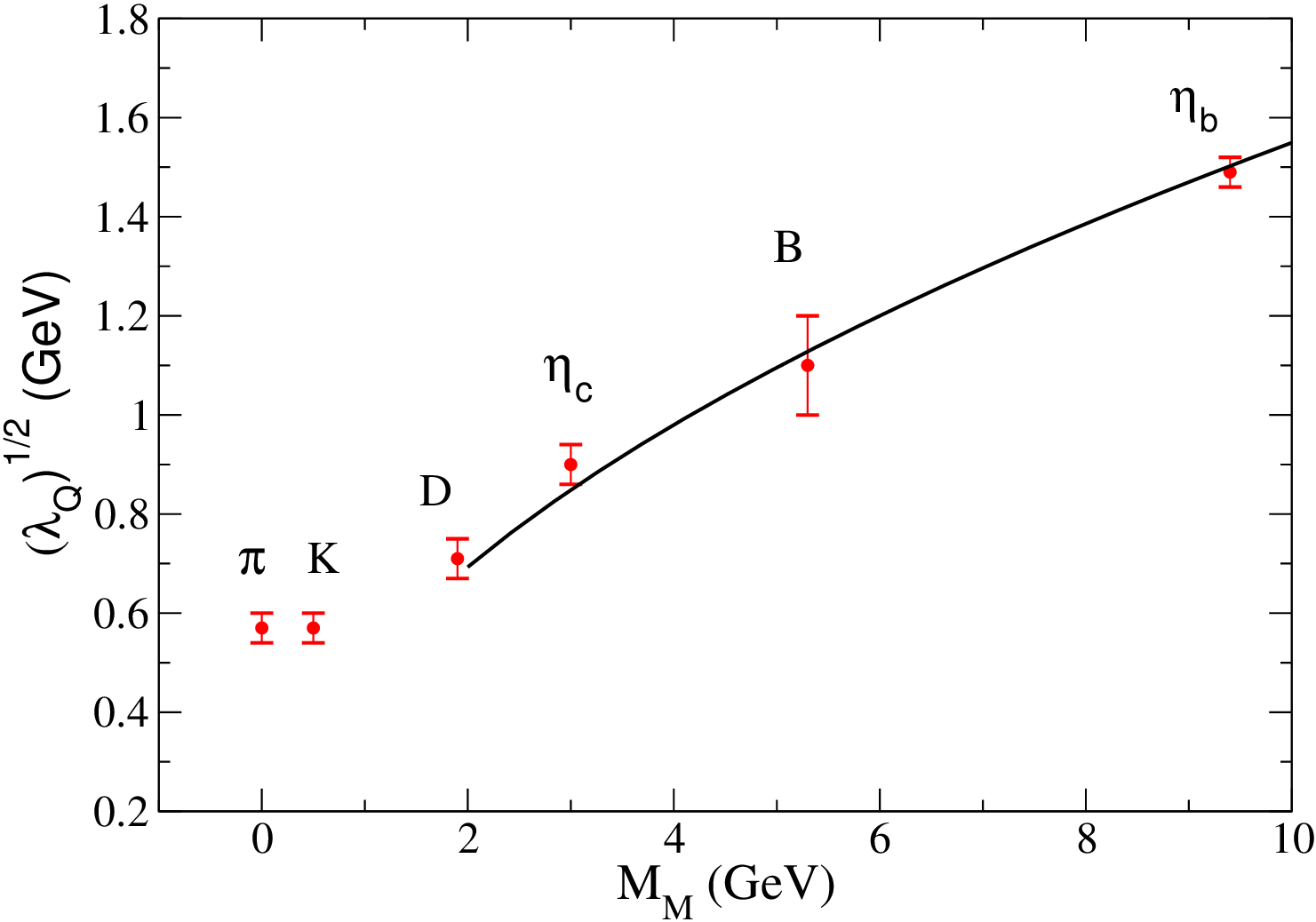}
%\vspace{10pt}
\caption{\label{lambda}  Fitted values of $\sqrt{\la_Q}$ as a function of the mass of the lowest meson state on the trajectory. The solid line is the fit from Eq.~\req{hq}.}
\end{figure}

In Fig. \ref{lambda} we show the values of $\lambda_Q$ for the
$\pi,\,K,\,D,\eta_c,\,B$ and $\eta_b$ meson families  as a function  of the
meson mass $M_M$. For the light quarks we are far away from the
heavy quark limit result \req{hq}. It is remarkable  that the
simple functional dependence \req{hq} derived in the heavy quark
limit works very well  for all heavy states, including double-heavy states. This shows universal behavior for all heavy states, including the double-heavy and the heavy-light states. In contrast, HQET is applicable only for states with only one heavy quark.
Fitting the results in Table~\ref{lamb-ta} for $M_M\geq 1.87$ GeV with Eq.~\req{hq} one finds
\beq \lb{Cnum} 
C=(0.49\pm0.02) \mbox{ GeV}^{1/2}. 
\enq
This value agrees, within the errors, with the value obtained in \cite{Dosch:2016zdv} from heavy-light hadrons.

\vspace{20pt}
%%%%%%%%%%%%%%%%%
%%%%%%%%%%%%%%%%

\subsection{ Excitation energies of heavy mesons}

We can use Eq.~\req{hq}  with \req{Cnum} to address a longstanding puzzle in the quarkonia spectrum \cite{Quigg:1979vr}: Why are the excitation energies of the heavy mesons approximately independent of the heavy quark mass?

From Eq.~\req{mesfin} one sees that $\Delta M^2[m_1,m_2]  \equiv M_0^2$ is the mass of the lowest meson state on the  trajectory with $n = L_M = S_M = 0$. Therefore one can write:
\beq
M_M^2[n,L_M,S_M]= M_0^2\left(1 + \frac{4 \la_Q}{M_0^2}(n+L_M+S_M/2)\right)=M_0^2\left(1 + \frac{4 C^2}{M_0}(n+L_M+S_M/2)\right),
\enq
where, from \req{hq} and \req{Cnum}, $C^2=\frac{\la_Q}{M_0}\sim0.24$ GeV. Thus, for heavy quark masses:
\beq
M_M[n,L_M,S_M] \approx M_0+ 2 C^2(n+L_M+S_M/2),
\enq
which implies
\beq\lb{exc}
M_M(1,L_M,S_M)-M_M(0,L_M,S_M)=2 C^2\sim 480\mbox{ MeV}.\
\enq

The prediction in \req{exc} shows that the excitation energies of the heavy mesons are indeed independent of the heavy quark mass. The experimental mass differences are
%\cite{PDG}:\\
% $\et_c(2S)-\et_c(1S):\;  655\mbox{ MeV}; \quad \ps_c(2S)-J/\ps(1S) :\; 589\mbox{ MeV};$\\
%$\et_b(2S)-\et_b(1S) :\; 601\mbox{ MeV}; \quad  \Up(2S)-\Up(1S) :\; 563\mbox{ MeV}, $\\ which are
consistent within the expected model uncertainties, as a first order approximation to the QCD theory. As a matter of fact, one would expect that gluon exchange would play an important role for small size states.

\subsection{Predictions for higher excitations of  charmonium and bottomonium}

%%%%%%%%%%%%%%%%%%%%%%
\begin{table}[h]
\begin{center}
\begin{tabular}{| ccc | ccc|ccc|}
\hline
\multicolumn{3}{|c|}{Meson} & \multicolumn{3}{c|}{Baryon} & \multicolumn{3}{c|}{Tetraquark}\\
$q$-cont&$~J^{P}$ & Name & $q$-cont & $J^{P}$ &Name &$q$-cont &$~~J^{P}$& Name \\
\hline
$\bar{b}c$&$0^-$&${B_c}(6275)$& --- &---&---& --- & --- &--- \\
$\bar{b}c$&$1^+$&${B}_{c1}(\sim6750)$& $[bq]c$ & $(1/2)^+$&$\Xi_{cb}(\sim6750)$ & $[bq][\bar{c}\bar{q}]$ & $0^{+}$ &${B}_{c0}(\sim6750)$ \\
\hline
\end{tabular}
\end{center}
\caption{\small \label{supar}
Quantum number assignment for the $B_c$ trajectory and baryonic and tetraquark superpartners.}
\end{table}
%%%%%%%%%%%%%%%%%%%%%%%

The relation in Eq.~\req{hq}  together with \req{Cnum} also allows us to determine the value: $\sqrt{\la_Q}=  (1.23\pm0.05)$ GeV, for the $B_c(6275)$, a $J^P=0^-$ state. Using this value of $\sqrt{\la_Q}$, we can also predict the masses for the mesons on the $B_c(6275)$ trajectory.
The prediction for the  $B_{c1}$ ($J^P=1^+$) mass is
\beq \lb{bc1} 
M_{B_{c1}}=(6.75\pm0.10) \mbox{ GeV}.
\enq
Therefore, from supersymmetry we predict a similar mass for its baryonic superpartner, $\Xi_{cb}$, and for the tetraquark superpartner, ${B}_{c0}$ ($J^P=0^+$). We show these predictions in  Table~\ref{supar}.
Our prediction for the $B_{c1}$ mass is in excellent agreement with the recent lattice  estimate: $M_{B_{c1}}=(6.726\pm0.016)$ GeV \cite{Mathur:2016hsm}.  For other model predictions for the $B_{c1}$ mass see, for instance, \cite{Kiselev:1994rc,Wang:2012kw}.

In Table~\ref{xibc} we show other model predictions for the baryonic superpartner of the $B_{c1}$  meson, the  $\Xi_{cb}$ state. Comparing the numbers in Table~\ref{xibc}, we can see that our predictions are in good agreement with most of the previous model predictions.

%%%%%%%%%%%%%%%
\begin{table}[h]
\begin{tabular}{|c|c|c|c|c|c|c|c|}
\hline  
 $M_{\Xi_{cb}}$(GeV)\,  & \,$6.86\pm0.28$\, & \,$6.933$\, & \,$6.75\pm0.05$\, & \,$6.72\pm0.20$\, &\, $6.92\pm0.13$\, &\, $6.835\pm0.015$\, &\, $6.75\pm0.10$\, \\
\hline 
Ref.  & \cite{Bagan:1994dy} & \cite{Ebert:2002ig}& \cite{Zhang:2008rt} & \cite{Aliev:2012ru} &   \cite{Karliner:2017elp} & \cite{Gershtein:2000nx} & this work\\
\hline 
\end{tabular} 
\caption{\label{xibc} Predictions for the $\Xi_{cb}$ mass.}
\end{table}

In our approach the $B_{c0}$  state is considered as the tetraquark superpartner of the baryon $\Xi_{cb}$. Other predictions for the mass of the $B_{c0}$ state, also  considered as four-quark state, are shown in the first three columns in Table~\ref{bc0}.
In the case of $B_{c0}$ our prediction is somewhat smaller than most of the previous predictions, but still in agreement within the errors. There is also a prediction, from lattice gauge theory, for the $B_{c0}$ mass. However, in the lattice calculation the $B_{c0}$ is considered as a $c\bar{b}$ state. The predicted mass is included in the last column in Table~\ref{bc0} and it is still in agreement with our prediction within the errors.

%%%%%%%%%%%%%%
\begin{table}[h]
\begin{tabular}{|c|c|c|c|c|c|}
\hline  
$M_{B_{c0}}$( GeV)\,  & \,$7.15\pm0.05$\, & \,$6.97\pm0.19$\, & \,$6.77\pm0.11$\, & \, $6.75\pm0.10$\, &\,$6.690\pm0.016$\,\\ \hline 
Ref.  & \cite{Chen:2013aba} & \cite{Agaev:2016dsg} & \cite{Albuquerque:2012rq} & this work & \cite{Mathur:2016hsm}\\
\hline 
\end{tabular} 
\caption{\label{bc0} Predictions for the $B_{c0}$ mass.}
\end{table}
%%%%%%%%%%%%%%%%%%

We can also use the information extracted from the fits in Figs.~\ref{charm} and \ref{beauty} to predict the masses of higher orbital excitations of the charmonium and bottomonium states. We show our predictions in Table~\ref{higher}, where we also show predictions based on different models.

%%%%%%%%%%%%%%%%%%
\begin{table}[ht]
\begin{tabular}{|c|c|c|c|c|c|c|}
\hline  
\multicolumn{7}{|c|}{charmonium}\\
\hline
state\,  & $^{2S+1}L_J$ \, & \, this work\, & \,\cite{Bhavsar:2018umj} &\, \cite{Shah:2012js} \, &\,\cite{Barnes:2005pb}\,&\,\cite{Li:2009zu}\,
\\ \hline 
$\eta_{c2}$ & $^1D_2$  & $3.90\pm0.09$ & $3.662$ & $3.802$&3.799 &3.796\\
$\psi_{3}$ & $^3D_3$  & $3.93\pm0.11$ & $3.770$ & $3.843$&3.806 & 3.799\\
\hline
\multicolumn{7}{|c|}{bottomonium}\\
\hline
state\,  & $^{2S+1}L_J$ \, & \, this work\, & \,\cite{Bhavsar:2018umj} &\,\cite{Shah:2012js}\,&\, \cite{Godfrey:2015dia}\,&\,\cite{Wang:2018rjg}\, \\
\hline 
$\eta_{b2}$  &  $^1D_2$ & $10.30\pm0.10$ &10.068 &10.166 &$10.148$& 10.163\\
$\Upsilon_{3}$ & $^3D_3$  & $10.32\pm0.09$ & 10.140&10.177& $10.155$ & 10.170 \\
\hline 
\end{tabular} 
\caption{\label{higher} Mass spectrum of  the predicted charmonium and bottomonium orbital excited states. All mass values are in GeV.}
\end{table}

\subsection{ Model uncertainties}

It should be noted that the errors quoted in our results in Tables \ref{xibc}, \ref{bc0} and \ref{higher} where obtained considering only the uncertainty in the value of $\lambda_Q$ and in the quark masses. These errors are probably underestimated, therefore our uncertainties should be considered as lower limits.

From Table~\ref{higher} we can see that our predictions, in the case of charmonia, are in agreement with most of the other predictions, considering the errors. It is very interesting to notice that our prediction for the mass of the $\Psi_3$ state is in excellent agreement with the mass of the first radial excitation of the $\chi_{c2}(3556)$: the $n=1$ $\chi_{c2}(3927)$ state, and in a good agreement (within the error) with the $n=2$ and $L_M=0$ $\psi(4039)$ state. According to Eq.~\req{mesfin}, states with $n=0$ and $L_M=2$ should have the same mass as the states with $n=1$ and $L_M=1$ or $n=2$ and $L_M=0$, if they have the same $S_M$, as the case of the  $\psi_3$, $\chi_{c2}(3927)$ and $\psi(4039)$ states respectively.

In the case of bottomonia our predictions are somewhat higher than previous model calculations. However, as in the case of $\Psi_3$ discussed above, our prediction for the mass of the $\Upsilon_3$ state is in a good agreement with the mass of the first radial excitation of the $\chi_{b2}(9910)$: the $n=1$ $\chi_{b2}(10270)$ state, and in excellent agreement with the $n=2$ and $L_M=0$ $\Upsilon(10355)$ state. Unfortunately, there are no other observed radial excited states ($n=1$) to be compared with the other predictions in Table~\ref{higher}. We urge the experimentalists to make an effort to measure the masses of the predicted states.
The two examples discussed above  show that, even in the case of the double-heavy states, the Regge slope is the same in both $n$ and $L$ quantum numbers, as predicted by the SuSyLFHQCD.

Since $\lambda_Q$ and $\Delta M^2$, in Eq.~\req{mesfin}, are determined from the fits to the Regge trajectories to the different meson families, we can use these values in \req{delm2} to estimate the effective heavy quark masses. We obtain $m_c=(1.52\pm0.07)$ GeV and $m_b=(4.63\pm0.04)$ GeV. Using the same procedure, the heavy quark masses obtained from heavy-light systems in \cite{Dosch:2016zdv}  were $m_c=1.55$ GeV and $m_b=4.922$ GeV, indicating the inherent uncertainties of the model.

\subsection{Decay constants}

The decay constant $f_M$ of a pseudoscalar  meson is the coupling of the hadron to its current. In a bound state model for mesons it is related to the value of the LF wave function at the origin~\cite{Brodsky:2007hb}.
 \beq \label{fm2} 
 f_M=\sqrt{\frac{2 N_C}{\pi}}\int_0^1dx\, \psi(x, \vec{b}_\perp = 0). 
 \enq
Using the result in \req{wf} one has for a meson with two constituents of equal mass $m_Q$:
 \beq \label{fm} 
 f_M=N_m{\sqrt{2 N_C\la_Q}\over\pi}\int_0^1dx\, \sqrt{x(1-x)}\,e^{-m_Q^2/(2\la_Q x(1-x))},
 \enq
with (see \req{norm}):
\beq\lb{norm2}
N_m^2={1\over\int_0^1dx~e^{-{m_Q^2/(\la_Q x(1-x))}}}.
\enq
 For the particular case of spin projection zero, the radiative vector meson decay constant is also given by Eq.~\req{fm} \cite{Branz:2010ub,Vega:2009zb}.

The values of the quark masses and $\la_Q$ determined in the previous section can be used to evaluate the meson decay constant in \req{fm}. In the limit of very large quark masses the integrals in \req{fm} and \req{norm2} can also be approximately evaluated analytically by the saddle point method, since in the limit $m_Q \to \infty$ the function  $e^{-m_Q^2/(2\la_Q x(1-x))}$ is very sharply peaked at $x=\half$.  Introducing $z^2= m_Q^2/\la_Q$, one obtains: ${N_m}= e^{2 z^2} \frac{2 \sqrt{z}}{\pi^{1/4}}$, and
\beq
\int_0^1dx\, \sqrt{x(1-x)}\,e^{-z^2/(2 x(1-x))}= {e^{-2 z^2}\over 2 z} \sqrt{\frac{\pi}{8}}.
\enq 
This leads to the asymptotic value of the decay constant: 
\beq
f_M= \sqrt{\frac{3}{4}} \pi^{- 3/4} \frac{\sqrt{ \la_Q}}{\sqrt{m_Q}} \, \la_Q^{1/4} \approx (0.178~\mbox{ GeV}^{3/4}) \,  M_M^{1/4} ,
\enq
where we have used $M_M\sim2~m_Q$ and $C= \frac{\sqrt{ \la_Q}}{\sqrt{M_M}} \approx 0.49 \; \mbox{ GeV}^{1/2}$, see \req{hq} and \req{Cnum}.
LFHQCD therefore predicts an increase in the decay constant with the meson mass.

%%%%%%%%%%%%%%%
\begin{table}[ht]
\bec
\begin{tabular}{|c|c|c|c|}
\hline  
Meson & $f_M$ [MeV] & $f_V^{(exp)}/e_V$ [MeV] & $\hat e_V$\\
\,  & \,this work \, & \,Ref.~\cite{PDG}  & \\
\hline 
$\rho$& 160&221 $\pm$  5 &$\frac{1}{\sqrt{2}}$\\
$\om$  & $160$ & $196\pm3$ & $\frac{1}{3 \sqrt{2}}$\\
$\ph$  & $161$ & $229 \pm4$ & $\frac{1}{3 } $\\
$J/\psi $  & $228$ & $416 \pm5$ &$ \frac{2}{3 }$ \\
$\Upsilon $  & $299$ & $715 \pm5$ & $\frac{1}{3 }$ \\
\hline 
\end{tabular} 
\enc
\caption{\label{de-co} Decay constants for different vector mesons, The value $f_M$ is determined from \req{fm}, the experimental values in the third column are obtained from the radiative constant by dividing through the  effective quark charge $\hat e_V$.}
 \end{table}
%%%%%%%%%%%%%%

The radiative decay constants of neutral vector mesons are, up to the charge factor and radiative corrections,  the general decay constants \cite{barb}.  In  Table \ref{de-co} we display the results of \req{fm} together with the observed values, obtained from the electromagnetic decay constant divided by the effective quark charge $\hat e_V$. To show the tendency we have also included the results for the light vector mesons.

As shown in Table~\ref{de-co}, the increase of the decay constant with meson mass is indeed observed, but the theoretical increase is much too slow.  The agreement of the coupling constants obtained here with the experimental values is poor, as in other similar approaches \cite{Branz:2010ub,Vega:2009zb}. This suggests that, despite the fact that the eigenvalues obtained in this SuSyLFHQCD are in good agreement with the experimental values, the wave functions are too simple to convey all the complexity of the quarkonium states. Even for the light mesons the decay constant comes out to small, as can be seen in Table~\ref{de-co}.  The increasing discrepancy suggests that special effects play a role in heavy quarkonia. A probable cause is the color-Coulomb attraction, since its effect increases with increasing mass and correspondingly smaller radius of the quarkonium.
% {This might suggest that higher Fock components, the first one being $ |Q \bar{Q} g\rangle $, should be included in the wave function  of heavy-heavy states.}

In \cite{Branz:2010ub} the LF wave functions were modifyed by introducing a phenomenological longitudinal term, with a new dimensional parameter which scales as $\sqrt{m_Q}$, while keeping the dilaton parameter fixed. However, such modification of the wave function did not improve the agreement of the decay constants with the experimental values. In \cite{Chang:2018aut,Chang:2017sdl,Ahmady:2018muv} the wave function proposed in \cite{Branz:2010ub} was modifyed by considering a helicity-dependent holographic wavefunction. With this modification a better agreement, of the decay constants with data, is obtained in the case of light and heavy-light  mesons. However, the authors of \cite{Chang:2018aut,Chang:2017sdl,Ahmady:2018muv} have not studied  heavy-heavy mesons. Improvement for the predictions of the light vector mesons decay constants can also be obtained by extending the model to include dynamical spin effects in the LF wave functions \cite{Forshaw:2012im,Ahmady:2016ujw}.
In \cite{Braga:2015jca} the quarkonium decay constants were evaluated  directly from the two point correlator function, calculated at some finite value, $z=z_0$, of the radial coordinate of AdS$_5$ space. This corresponds to introducing a new energy scale: $1/z_0$, in the model that leads to a better agreement with data. We expect to further investigate the origin of this discrepancy.

\section{Summary and Conclusions \lb{CONCL}}

In this paper we have described the consequences of extending semiclassical light-front bound-state equations to double-heavy quark systems. The approach is based on supersymmetric light front holographic QCD.   The supersymmetry relates wave functions of mesons to baryons and of baryons to tetraquarks:  this approach is not based on a supersymmetric quantum field theory but on supersymmetric quantum mechanics~\cite{Witten:1981nf}.

We have shown that  the mass spectra of double-charmed and double-beauty mesons are compatible with  the linear Regge trajectories given in Eq.~\req{mesfin}. In particular, the remarkable equality of the Regge slopes in both, orbital angular momentum $L$ and principal quantum number $n$ is predicted to remain valid, even for hadrons containing double-heavy quarks. From the determination of the Regge slope from these trajectories, we have shown that this parameter follows the same relation \req{hq} as obtained in heavy-light systems from heavy quark symmetry~\cite{Dosch:2016zdv}. This relation  also explains an old puzzle in quarkonia physics: why the excitation energies are approximately independent of the heavy meson mass~\cite{Quigg:1979vr}. The relation in \req{hq} allowed us to make predictions for several double-heavy states, shown in Eq.~\req{bc1} and in Tables~\ref{xibc}, \ref{bc0} and \ref{higher}. Our predictions are in good agreement with other model predictions  in the case of charmonium, and in a fair agreement in the case of bottomonium.

We have also evaluated the radiative decay constant of the vector states $J/\psi$ and $\Upsilon$. The poor agreement with the experimental values of the decay constants shows that, although the eigenvalues obtained in this SuSyLFHQCD are in good agreement with the experimental values, the wave functions used are too simple to express all the complexity of the quarkonium states.

We have shown how supersymmetry, together with light-front holography, leads to connections between double-heavy mesons, baryons and tetraquarks, thus providing new perspectives for hadron spectroscopy and QCD.   We emphasize that measurements of additional states in the double-heavy quarks sector will test our predictions.

\begin{acknowledgments}

S.J.B. is supported by the Department of Energy, contract DE--AC02--76SF00515.  M.N. is supported by FAPESP process\# 2017/07278-5. SLAC-PUB-17259.

\end{acknowledgments}

%\newpage

\end{document}